\documentstyle[pra,multicol,aps,epsf]{revtex}
\def \I{{\rm i}}

\begin{document}

\title{Limits of noise squeezing in Kerr effect}

\author{Ji\v r\'\i\ Bajer,$^{1}$ Adam Miranowicz,$^{2}$ and Ryszard Tana\'s$^{2}$}

\address{$^{1}$Department of Optics, Palack\'y University,
17.~listopadu 50, 772~00 Olomouc, Czech Republic}
\address{$^{2}$Nonlinear Optics Division, Institute of Physics,
Adam Mickiewicz University, 61-614 Pozna\'n, Poland}

\date{published in {\em Czech. J. Phys.} {\bf 52} (2002) 1313-1319}
\maketitle

\widetext

\begin{abstract}

It is well known that the optical Kerr effect can be a source of
highly squeezed light, however the analytical limit of the noise
suppression has not been found yet. The process is reconsidered
and an analytical estimation of the optimal quadrature noise level
is presented. The validity of the new scaling law is checked
numerically and analytically.
\\ \\
{\em PACS:} 42.50.Ar, 45.65.Hw, 42.50.Dv\\
{\em Key words:} Kerr effect, anharmonic oscillator, squeezing,
photon statistics
\end{abstract}

\widetext
\begin{multicols}{2}

\section{Introduction}

The cubic non-linearity of isotropic media, like liquid or fiber,
can be responsible for the self-phase modulation, self-induced
rotation of polarization, self-modulation of intensity profile or
self-focusing of strong optical beam. These phenomena can simply
be explained by a linear dependence of the refraction index on the
intensity of the propagating light, which is usually referred to
as the {\em optical Kerr effect}. This effect plays also an
important role in, e.g., the non-diffracting beam technics,
soliton propagation or non-demolition measurements. From the
viewpoint of quantum optics, the Kerr effect can be a source of
the squeezed light
\cite{tanas83,yamamoto,milburn2,milburn1,gerry,luks1,buzek,tanas91,meng,sundar,shapiro}
(for a review see Ref. \cite{tanasrev}), also referred to as the
{\em self-squeezed} light. Unfortunately, the Kerr non-linearity
is usually very small, thus the effective application of the Kerr
effect requires the long-interaction times (or lengths) and
high-intensity lasers. The simplest strategy for the effective
Kerr process is to use a long optical fiber and strong laser
pulses. In recent experiments with the Kerr fibers, the quantum
noise was successfully reduced by $0.7$ dB \cite{levenson} or even
by $3.5$ dB \cite{haus} (for a discussion of some experimental
results see \cite{milburn2,bachor})

Quantum description of the Kerr-type evolution is among a few
non-trivial quantum dynamical models, which are fully solvable.
This is one of the reasons for its long-term popularity and
fundamental importance. Mathematically, the Kerr process can
simply be described by the well-known interaction Hamiltonian
\begin{equation}
\hat{H}=\frac{1}{2}\hbar g\hat{a}^{\dag
2}\hat{a}^{2}=\frac{1}{2}\hbar g\hat{n}\left( \hat{n}-1\right) ,
\label{N01}
\end{equation}
where $\hat{a}$ and $\hat{a}^{\dag}$ are, respectively, the
annihilation and creation operators satisfying the boson
commutation relation; $\hbar $ is the Planck constant and $g$ is
the Kerr non-linearity. Since the photon-number operator
$\hat{n}=\hat{a}^{\dag }\hat{a}$ commutes with the Hamiltonian,
the photocount statistics of the field propagating through the
Kerr medium is conserved. Nevertheless, the photon number
fluctuations can be reduced via the Kerr effect, e.g., using the
nonlinear Mach-Zehnder interferometer with the Kerr medium in one
of the arms \cite{yamamoto}.

Assuming the input field to be in coherent state $|\alpha\rangle$,
the solution of the Schr\"odinger equation is the Kerr state
\cite{tanasrev}
\begin{eqnarray}
|\psi _{K}\rangle &=& \exp \left[
-\I\frac{\tau}{2}\hat{n}(\hat{n}-1) \right] |\alpha\rangle
\nonumber\\ &=&e^{-|\alpha| ^{2}/2} \sum_{n=0}^{\infty}
\frac{\alpha ^{n}}{ \sqrt{n!}}\exp \left[ -\I\frac{\tau}{2}n\left(
n -1\right) \right] |n\rangle, \label{N02}
\end{eqnarray}
where $\tau=gL/v_g$ is the dimensionless rescaled interaction
length of the Kerr medium or the interaction time; $v_g$ is the
group velocity of light in the medium. The only and trivial
difference between the spatial and temporal evolution is the
opposite sign in front of $\tau$. Thus, we use the terms
interaction time or length interchangeably.

\section{Noise squeezing analysis}

The Kerr dynamics has been thoroughly studied from different
points of view. Many results predicted by these theoretical
studies are well known. For example, it can be proved that the
time evolution is exactly periodic with the period $2\pi $. At the
time moments $\tau _{N}=2\pi M/N$, being rational fractions of the
period, the input coherent state is transformed into a
superposition of two \cite{yurke} or more \cite{miranowicz}
well-localized coherent states, referred to as the Schr\"odinger
cats and kittens, respectively. For the present study, the most
important result is the fact that the Kerr states evolve
periodically into the quadrature noise squeezed states
\cite{tanas83,milburn1,gerry,luks1,buzek,tanas91,meng,sundar}. By
defining the general quadrature component as $\hat{X}_{\theta
}=\hat{a}\exp \left( \I\theta \right) +\hat{a}^{\dag }\exp \left(
-\I\theta \right)$, where $\theta$ represents the controlled phase
of a homodyne detector, the principal squeezing can be defined as
the minimum of all the quadrature variances with respect to all
the possible phases $\theta$ and the following relation holds
\cite{luks1,luks2}
\begin{eqnarray}
S &=& \min_{\theta }{\rm Var} \hat{X}_{\theta}
\nonumber \\
&=& 1+2\left( \langle \hat{a}^{\dag }\hat{a}\rangle -|\langle
\hat{a}\rangle|^2 \right) -2\left| \langle \hat{a}^{2}\rangle
-\langle \hat{a}\rangle ^{2}\right| . \label{N03}
\end{eqnarray}
\hspace*{30mm} (a) \hspace*{35mm} (b)
\begin{figure}
\vspace*{-5mm} \hspace*{-4mm} \epsfxsize=4.1cm \epsfbox{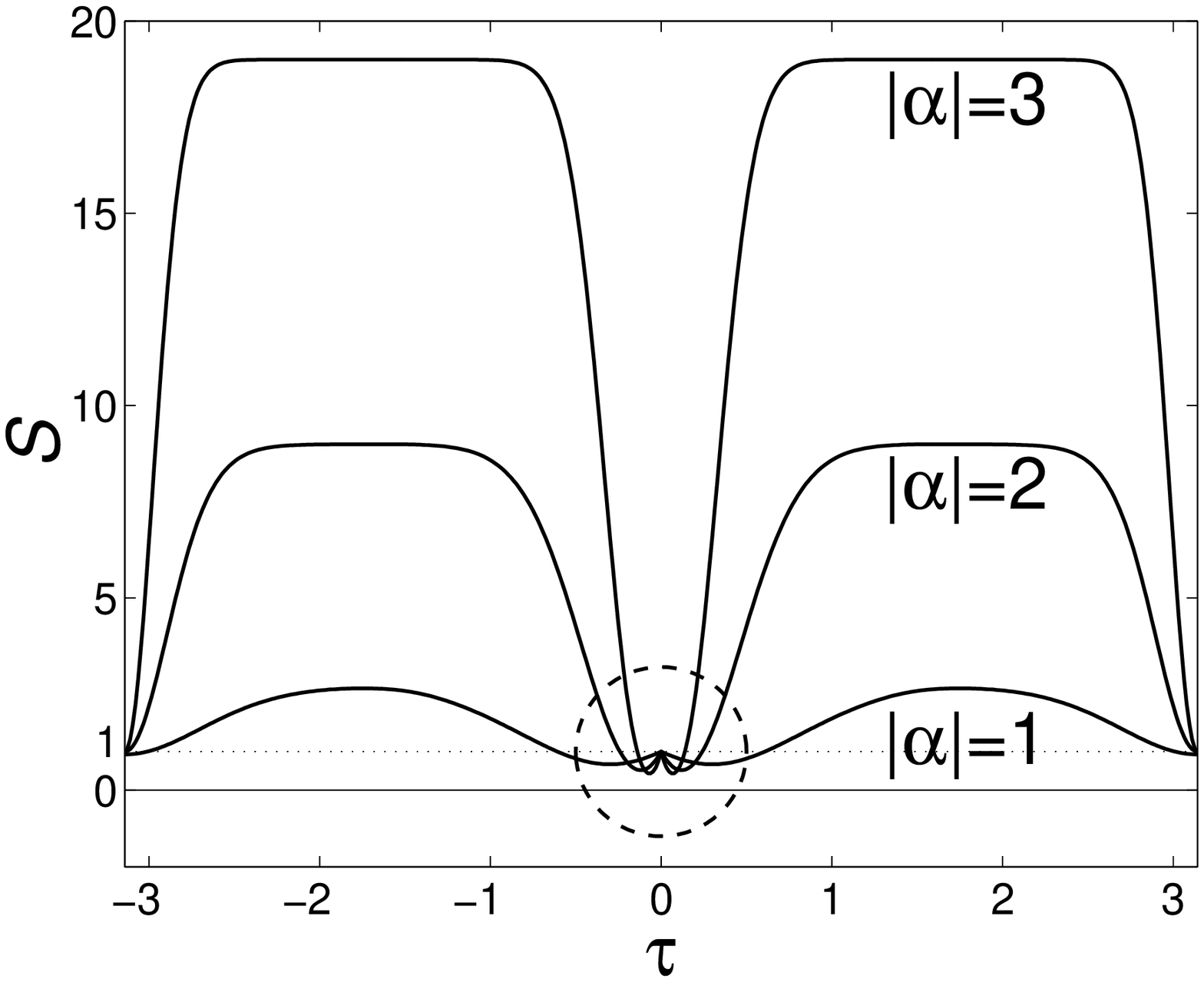}
\hspace*{1mm}\epsfxsize=4.25cm \epsfbox{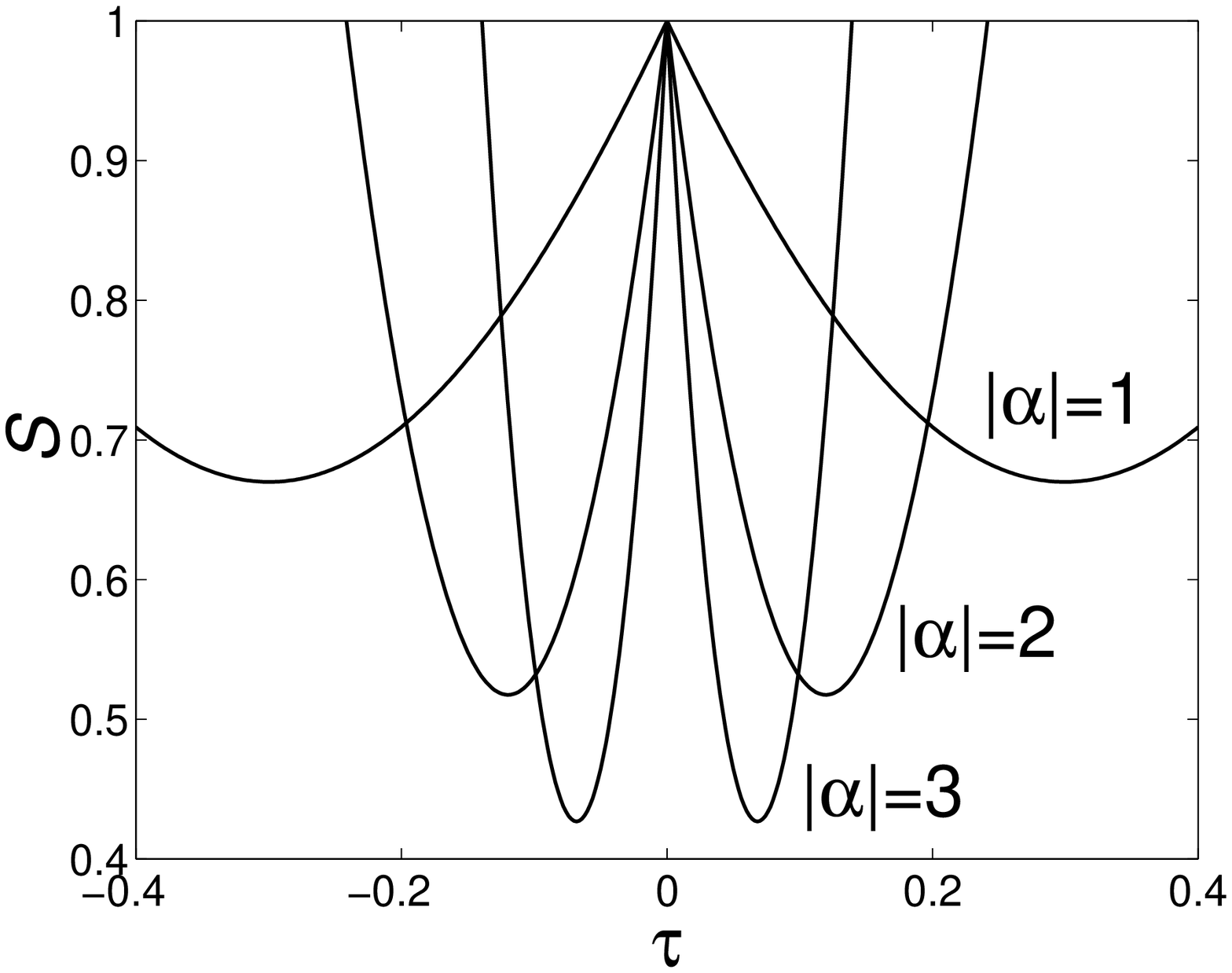}
\vspace*{0mm}%
\caption{Time dependence of the principal squeezing $S$ for
several values of the input amplitudes $|\alpha|$. Dotted line
represents the vacuum noise level of $S=1$. We are particularly
interested in the region of the suppressed quadrature noise, which
is marked in the circle in figure (a) and presented enlarged in
figure (b).
} \end{figure}%
Geometrically, the principal squeezing represents the smaller
half-axis of Booth's elliptical lemniscate \cite{loudon}. For
vacuum and coherent fields the quadrature noise is totally
independent of the choice of phase $\theta $ and it holds $S=1$.
This level of vacuum noise represents the quantum noise limit.
Squeezing of the quadrature noise below the vacuum noise level
occurs if $S<1$. The formula for the principal squeezing in the
Kerr effect follows from solution (\ref{N02}) and reads as
\begin{eqnarray}
S &=&1+2| \alpha | ^{2}\left\{ 1-\exp \left[ 2| \alpha
| ^{2}\left( \cos \tau -1\right) \right] \right\}  \nonumber \\
&&-2| \alpha | ^{2} \Big| \exp \left[ | \alpha | ^{2}\left( e^{ -2
\I \tau} -1 \right) -\I\tau \right]
\nonumber \\
 &&-\exp \left[ 2 | \alpha | ^{2}\left( e^{
-\I \tau} -1 \right) \right] \Big| . \label{N04}
\end{eqnarray}
The time dependence of $S$ for some amplitudes is presented in
figure 1. As for the Kerr state, the evolution of the principal
squeezing is periodic with the period of $2\pi$. Moreover, $S$ is
an even function, therefore it is sufficient to study the
evolution of $\tau$ in the interval of $\left( 0,\pi \right)$
representing a half of the whole period. From figure 1 we can see
that the principal squeezing (\ref{N04}) starts from the vacuum
noise level $ S\left( 0\right) =1$ and then evolves into the
squeezed states with $S<1$ for some limited time interval. At the
moment $\tau _{\min }$ the quantum noise is maximally squeezed up
to the value $S_{\min }=S(\tau _{\min})$. The phase portrait of
the state has a crescent shape. The noise level can be very highly
squeezed, especially for the large input amplitude $|\alpha|$, but
it never reaches the absolute zero. In the following, we will give
an analytical estimation of the maximum possible noise squeezing.
For larger interaction times, the quantum noise of the Kerr state
goes over from the vacuum noise level with $S=1$ up to the value
of $ S\approx 1+2| \alpha | ^{2}$. That maximum is very flat (see
figure 1(a)) and not particularly interesting for our purpose. The
corresponding phase portrait rapidly changes from the crescent
shape to the ring or to a discrete superposition of coherent
states. At the time $\tau =\pi /2$, the Kerr state is an
equally-weighted superposition of four coherent states. At last,
close to $\tau \approx \pi ,$ the quantum noise drops down to the
vacuum noise level $S\approx 1$. The phase portrait at this moment
corresponds to an equally-weighted superposition of two coherent
states.

The Kerr non-linearity is usually very small and, practically, the
interaction times of the order of $\tau \approx 10^{-6}$
\cite{ritze,tanasrev} can be reached in the optical domain.
Fortunately, as follows from the above analysis and presented in
figures 1(b) and 2, the highest noise suppression can be expected
for short interaction times. It is worth noting that much higher
Kerr-type non-linearities are observed in atom optics in, e.g.,
schemes based on Raman laser excitations of a trapped atom
\cite{wallentowitz}.

The noise power (\ref{N04}) depends on the interaction time $\tau$
and the input coherent amplitude $|\alpha|$. Instead of the
amplitude, the Kerr parameter $r=| \alpha | ^{2}\tau $ is often
used as a good measure of the Kerr interaction. For intense laser
pulses, the Kerr parameter can reach the values of the order of
unity \cite{bachor}. In real experiments, the interaction time is
fixed by the optical fiber length, but the optical power, and thus
the Kerr parameter, can simply be controlled. In the following, we
will study the principal squeezing $S\left(r\right)$ in its
dependence on the Kerr parameter for the fixed interaction times.

From the point of view of potential applications, it is
interesting to know the level of the minimum noise, which can be
suppressed in the Kerr process. Equation (\ref{N04}) is tricky to
find the minimum of quantum noise level analytically. In the
former studies, it was solved only numerically and analyzed
graphically. We have plotted the minima of the noise level for
some values of the scaled interaction times in figure 2. The
regular dependence of the noise minimum on the interaction time is
evident. From the linear dependence of the minimum points on the
logarithmic scale we can guess the scaling law of the form
$S_{\min }\sim r_{\min }^{\gamma}$, which will be derived in next
section.

It is even hard to analyze numerically equation (\ref{N04}). The
trouble comes from computations that require extra computer
precision. For example, the curves in figure 2 for parameters
$\tau <10^{-6}$ cannot be computed directly from (\ref{N04}) using
the standard 16-digit arithmetics. The correct computation
requires a special computing system, like MAPLE or MATHEMATICA,
that enables at least the 100-digit arithmetics. On the other
hand, the same numerical problems can sometimes be overcome by
applying the saddle-point technique \cite{tanas91}.

\section{Noise squeezing approximation}

Here, as the main result of this Letter, we will derive an
approximation of (\ref{N04}) for large amplitudes and short
interaction times. This approximation can be used for simple
calculation of the noise power bypassing the computer rounding
error and will be used further for finding an analytic estimation
of the quadrature noise minimum.

First, we exclude the amplitude from (\ref{N04}) by the
substituting $|\alpha| ^{2}=r/\tau$. Assuming that the interaction
time $\tau $ is very short, we can expand the terms $\cos \tau
-1$, $\exp(-2 \I \tau) -1$ and $\exp(-\I \tau) -1$ in Taylor
series. To derive a useful approximation, we assume the Kerr
parameter \linebreak
\begin{figure}
\vspace*{-3mm} \hspace*{0mm} \epsfxsize=7.5cm \epsfbox{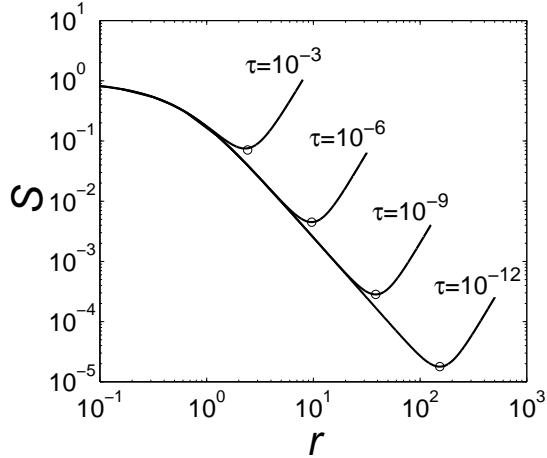}
\vspace*{2mm}%
\caption{Comparison of the principal squeezing minima as a
function of $r=|\alpha|^2\tau$ for several values of $\tau$. The
minima apparently lie on the straight line on the log-log scale.
The circles represent $S'_{\min}$ calculated from the scaling law
formula (\ref{N07}). Note a good accuracy of our estimation of the
minimum noise by $S'_{\min}$ for $\tau<10^{-3}$.
} \end{figure}%

\hspace*{32mm} (a) \hspace*{37mm} (b)
\begin{figure}
\vspace*{-5mm} \hspace*{-5mm} \epsfxsize=4.4cm \epsfbox{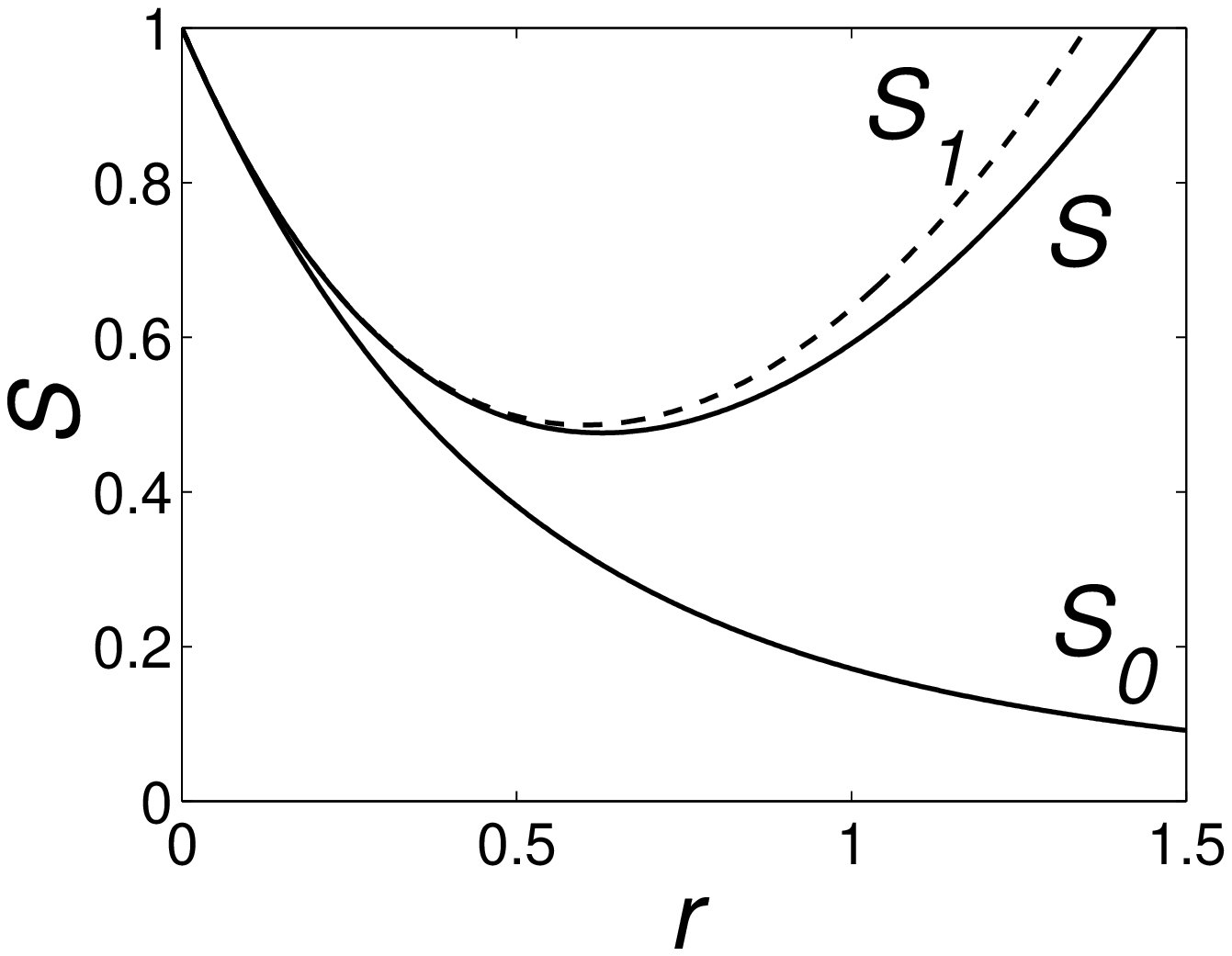}
\hspace*{0mm}\epsfxsize=4.3cm \epsfbox{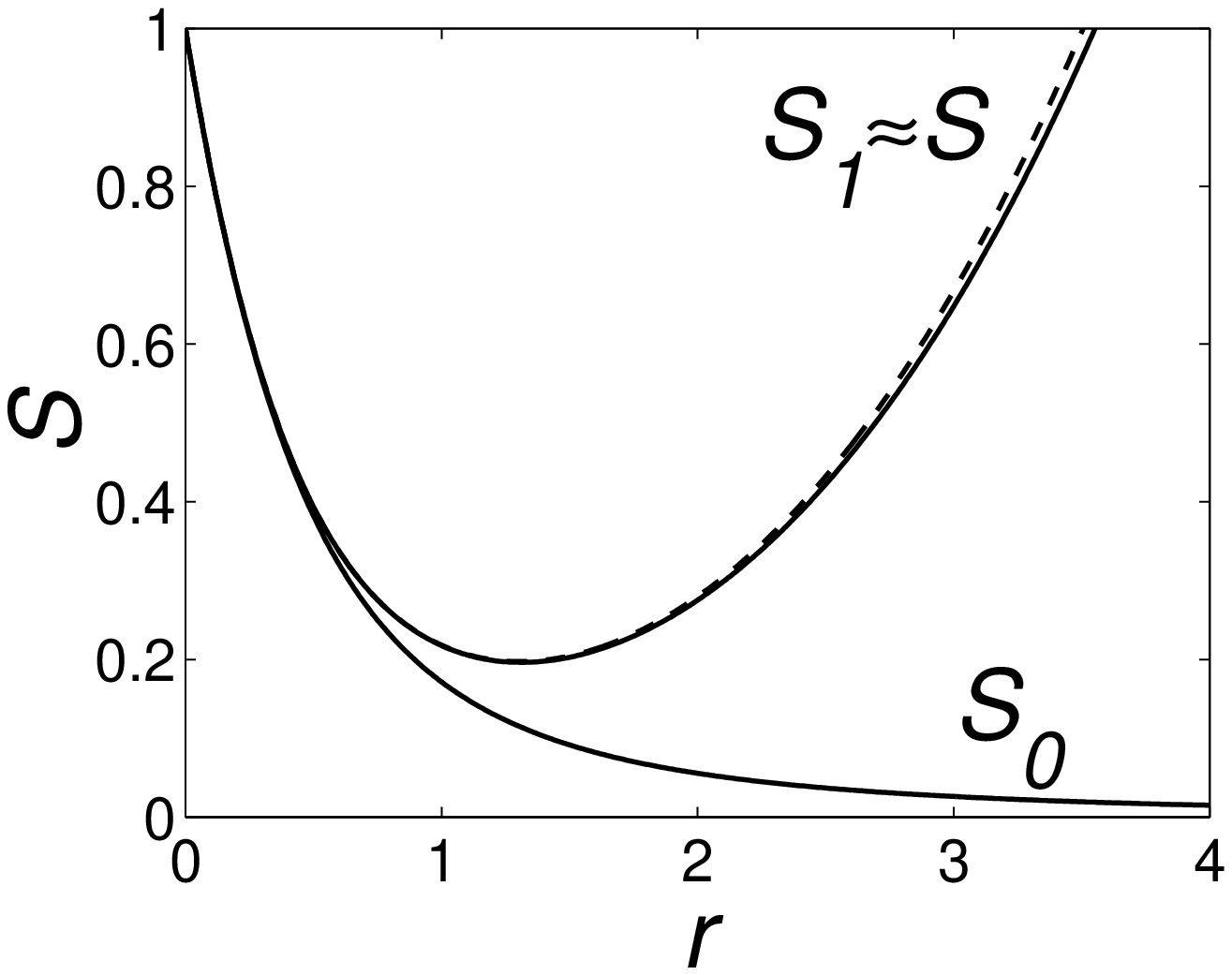}
\vspace*{0mm}%
\caption{Comparison of the approximations $S_{0}$ and $S_{1}$, and
the exact values $S$ of the principal squeezing as a function of
the Kerr parameter $r$ for (a) $\tau=10^{-1}$ and (b)
$\tau=10^{-2}$. The figure shows that, for $\tau\le 10^{-2}$, the
approximation $S_{1}$ can be used to estimate the minimum
quadrature noise with a good precision.}
\end{figure}%
\noindent $r$ to be finite and we sort all the terms in powers of
$\tau$. The Taylor expansion of all exponential functions in
(\ref{N04}) leads, after final arrangement, to the approximation
\begin{equation}
S_{1}=1+2r^{2}-2r\sqrt{1+r^{2}}-r^{3}\tau +\frac{\left(
3r^{2}+5\right) r^{2}\tau }{\sqrt{1+r^{2}}}, \label{N05}
\end{equation}
which is linear in $\tau$. The accuracy of the approximation
(\ref{N05}) in comparison to the exact values of squeezing for two
different values of $\tau$ can be seen in figure 3. In the less
accurate approximation, the linear correction in (\ref{N05}) can
be ignored. In that case we get the well-known approximation
\cite{tanas91,bachor}
\begin{equation}
S_{0}=1+2r^{2}-2r\sqrt{1+r^{2}},  \label{N06}
\end{equation}
which is indeed not sufficient to determine the minimum noise
level. Equation (\ref{N06}) depends on the Kerr parameter $r$ only
and thus gives physical grounds for the introduction of $r$. As
shown in figure 3, the $S_{0}$ approximation is monotonic in $r$
and has the only minimum of $S_{0}=0$ in the limit of
$r\rightarrow \infty $. For sufficiently large $r$, $ S_{0}$ can
be estimated as $S_{0}\approx 1/(4r^{2})$. On the other hand, for
the relatively small Kerr parameters, (\ref{N06}) can be
approximated linearly as $S_{0}\approx 1-2r$. The vast majority of
the up-to-date measurements have been performed in the domain
where the approximation (\ref{N06}) is valid.

By assuming that the Kerr parameter of the minimum searched is
large (see the values in Table 1), we can even more simplify
(\ref{N05}). We expand (\ref{N05}) in the asymptotic series for
large $r$ and keep the largest two terms only. Thus, we obtain
\begin{equation}
S_1 \approx S' =  \frac{1}{4r^{2}}+2r^{3}\tau .
\end{equation}
From this elementary approximation we obtain directly a formula
for the minimum of the principal squeezing
\begin{equation}
S'_{\min}= \frac{5}{12}\left( 12\tau \right) ^{2/5}\quad {\rm
at}\quad r'_{\min }= \left( 12\tau \right) ^{-1/5} \label{N07}
\end{equation}
leading to $S'_{\min}=5/(12 r'^{2}_{\min})$. This estimation of
the minimum noise level is an analytical confirmation of the
scaling law as observed in figure 2. We have shown that the
minimum level of quadrature noise is proportional to the $2/5$th
power of the interaction time or the fiber length. The shorter
interaction time $\tau$ (mod$2\pi$) results in the deeper
reduction of noise, but the higher optical power is required to
reach the minimum since it holds $|\alpha | ^{2}=r_{\min }/\tau
\sim \tau ^{-6/5}$. For example, to obtain $S_{\min }\approx
4.\,5\times 10^{-3}$ at the interaction time $\tau \approx
10^{-6}$, the optimum values of the Kerr parameter and the
intensity are $r_{\min }\approx 9.\,6$ and $| \alpha | ^{2}\approx
10^{7}$, respectively. A comparison of the exact values $S$ and
$\tau$, calculated from (\ref{N04}), with their estimations $S'$
and $\tau'$, given by the scaling law (\ref{N07}), is presented in
table 1 and figure 2, where the estimations are marked by small
circles.

\vspace{4mm}
\begin{table}
\caption{Several numerical values of the principal squeezing
minimum, $S_{\min}$, and the corresponding Kerr parameter
$r_{\min}$ obtained at the scaled interaction times $\tau$. Their
estimations $S'_{\min}$ and $r'_{\min}$ are based on the scaling
law approximation given by (\ref{N07}). It can be concluded from
the table that (i) the level of the minimum quadrature noise
regularly drops down with the shorter interaction time $\tau$
which, however, requires higher intensity such that the Kerr
parameter $r=|\alpha|^2\tau$ increases, and (ii) the estimation is
valid for $\tau<10^{-3}$.}
\begin{center}
\begin{tabular}{ccccc}
 $\tau $ & $S_{\min}$ & $S'_{\min}$ & $r_{\min }$ & $r'_{\min}$ \\
 \hline $10^{-1}$ & $0.\,476$ & $0.\,448\,$ & $0.\,627$  & $0.\,964\,$ \\
 $10^{-2}$ & $0.\,196$ & $0.\,178\,$ & $1.\,312$ & $1.\,528$ \\
 $10^{-3}$ & $7.\,503\times 10^{-2}$ & $7.\,103\times 10^{-2}$  & $2.\,288$ & $2.\,421$ \\
 $10^{-4}$ & $2.\,899\times 10^{-2}$ & $2.\,828\times 10^{-2}$  & $3.\,755$ & $3.\,839$ \\
 $10^{-6}$ & $4.\,501\times 10^{-3}$ & $4.\,482\times 10^{-3}$  & $9.\,609$ & $ 9.\,642$ \\
 $10^{-9}$ & $2.\,829\times 10^{-4}$ & $2.\,828\times 10^{-4}$ & $38.\,38$  & $ 38.\,38$ \\
 $10^{-12}$  & $1.\,784\times 10^{-5}$ & $1.\,784\times 10^{-5}$ & $152.\,8$ & $ 152.\,8$
\end{tabular}
\end{center}
\end{table}

\section{Conclusion}

We have analyzed ordinary coherent light interacting with a
non-absorbing non-linear Kerr medium, modelled as an anharmonic
oscillator. Former numerical studies have shown that the model
leads to almost complete quadrature squeezing. We have found, as
we believe for the first time, an approximate analytical formula
for the optimal noise level for high intensity coherent light
assuming realistic values (see, e.g., \cite{ritze}) of the Kerr
nonlinearity.

\vspace{3mm} \noindent JB was supported by the EU grant under the
QIPC project IST-1999-13071 (QUICOV) and the Czech Ministry of
Education within the project LN00A015. AM was supported by the
Japan Science and Technology Corporation (JST-CREST)


\end{multicols}

\begin{center}
\framebox{\parbox{0.75\columnwidth}{%
\begin{center}
published in {\em Czech. J. Phys.} {\bf 52}, No. 12 (2002) pp.
1313-1319
\end{center}}}
\end{center}

\end{document}